\documentclass[twocolumn]{aastex631}

\usepackage{natbib}
\usepackage{bm}
\usepackage{amsmath}
\usepackage{hyperref}
\usepackage{savesym}
\savesymbol{tablenum}
\usepackage{siunitx}
\usepackage{placeins}
\restoresymbol{SIX}{tablenum}
 \hypersetup{
     colorlinks=true,
     linkcolor=blue,
     filecolor=blue,
     citecolor = magenta,      
     urlcolor=red,
     }

\newcommand{\kms}{\textrm{km s}^{-1}}
\newcommand{\hi}{\text{H\,\sc{i}}}
\newcommand{\Msol}{\textrm{M}_{\odot}}
\newcommand{\QUaffiliation}{Department of Physics, Engineering Physics, and Astronomy, Queen’s University, Kingston, ON, K7L 3N6, Canada}
\newcommand{\ICRARaffiliation}{International Centre for Radio Astronomy Research (ICRAR), University of Western Australia, 35 Stirling Hwy, Crawley, WA 6009, Australia}

\newcommand{\sofia}{\textsc{SoFiA-2}}

\shorttitle{Comparing WALLABY to SIMBA}
\shortauthors{Perron-Cormier, M. et al.}
\begin{document}
\title{WALLABY Pilot Survey \& ASymba: Comparing $\hi$ Detection Asymmetries to the SIMBA Simulation}

\author{Mathieu Perron-Cormier}
\affiliation{\QUaffiliation}
\email{matpcormier@gmail.com}
\author{Nathan Deg}
\affiliation{\QUaffiliation}
\author{Kristine Spekkens}
\affiliation{\QUaffiliation}
\author{Mark L. A. Richardson}
\affiliation{\QUaffiliation}
\affiliation{Arthur B. McDonald Canadian Astroparticle Physics Research Institute, 64 Bader Lane, Kingston, ON, Canada,K7L 3N6}
\author{Marcin Glowacki}
\affiliation{
Institute for Astronomy, University of Edinburgh, Royal Observatory, Edinburgh, EH9 3HJ, United Kingdom}
\affiliation{
International Centre for Radio Astronomy Research, Curtin University, Bentley, WA 6102, Australia}
\affiliation{
Inter-University Institute for Data Intensive Astronomy, Department of Astronomy, University of Cape Town, Cape Town, South Africa}
\author{Kyle A. Oman}
\affiliation{Institute for Computational Cosmology, Physics Department, Durham University, South Road, Durham DH1 3LE, United Kingdom}
\affiliation{Centre for Extragalactic Astronomy, Physics Department, Durham University, South Road, Durham DH1 3LE, United Kingdom}
\author{Marc A. W. Verheijen}
\affiliation{Kapteyn Astronomical Institute, University of Groningen, Landleven 12, 9747 AD Groningen, the Netherlands}
\author{Nadine A. N. Hank}
\affiliation{Kapteyn Astronomical Institute, University of Groningen, Landleven 12, 9747 AD Groningen, the Netherlands}
\author{Sarah Blyth}
\affiliation{Department of Astronomy, University of Cape Town, Private Bag X3, Rondebosch 7701, South Africa}
\author{Helga Dénes}
\affiliation{School of Physical Sciences and Nanotechnology, Yachay Tech University, Hacienda San José S/N, 100119, Urcuquí, Ecuador }
\author{Jonghwan Rhee}
\affiliation{\ICRARaffiliation}
\author{Ahmed Elagali}
\affiliation{School of Biological Sciences, The University of Western Australia, Crawley, Western Australia, Australia}
\author{Austin Xiaofan Shen}
\affiliation{CSIRO, Space and Astronomy, PO Box 1130, Bentley, WA 6102, Australia}
\author{Wasim Raja}
\affiliation{CSIRO, Space and Astronomy, PO Box 1130, Bentley, WA 6102, Australia}
\author{Karen Lee-Waddell}
\affiliation{Australian SKA Regional Centre (AusSRC) - The University of Western Australia, 35 Stirling Highway, Crawley WA 6009, Australia}
\affiliation{CSIRO, Space and Astronomy, PO Box 1130, Bentley, WA 6102, Australia}
\affiliation{\ICRARaffiliation}
\author{Luca Cortese}
\affiliation{\ICRARaffiliation}
\author{Barbara Catinella}
\affiliation{\ICRARaffiliation}
\affiliation{Australia. ARC Centre of Excellence for All Sky Astrophysics in 3 Dimensions (ASTRO 3D), Australia.}
\author{Tobias Westmeier}
\affiliation{\ICRARaffiliation}

\begin{abstract}
An avenue for understanding cosmological galaxy formation is to compare morphometric parameters in observations and simulations of galaxy assembly.
In this second paper of the ASymba: Asymmetries of \hi\ in \textsc{Simba} Galaxies series, we measure atomic gas (\hi) asymmetries in spatially-resolved detections from the untargetted WALLABY survey, and compare them to realizations of WALLABY-like mock samples from the SIMBA cosmological simulations.  We develop a Scanline Tracing method to create mock galaxy $\hi$ datacubes which minimizes shot noise along the spectral dimension compared to particle-based methods, and therefore spurious asymmetry contributions. We compute 1D and 3D asymmetries for spatially-resolved WALLABY Pilot Survey detections, and find that the highest 3D asymmetries ($A_{3\mathrm{D}}\gtrsim 0.5$) stem from interacting systems or detections with strong bridges or tails.  We then construct a series of WALLABY-like mock realizations drawn from the SIMBA 50 Mpc simulation volume, and compare their asymmetry distributions. We find that the incidence of high $A_{3\mathrm{D}}$ detections is higher in WALLABY than in the SIMBA mocks, but that difference is not statistically significant ($p$-value = 0.05). The statistical power of quantitative comparisons of asymmetries such as the one presented here will improve as the WALLABY survey progresses, and as simulation volumes and resolutions increase.  
\end{abstract}

\section{Introduction}

The gas distributions of galaxies formed in a cosmological simulation can be compared to observations to constrain the galaxy formation model and determine the underlying drivers of galaxy morphology. These comparisons allow explorations of how this morphology is influenced by cosmological parameters, local environment effects, sub-grid physics, and more.  

Effective comparisons between observations and simulations require robust methods of creating mock observations from the simulated particle distributions. In atomic gas (\hi), the Mock Array Radio Telescope Interferometry of the Neutral ISM (MARTINI) code \citep{MARTINI_soft, MARTINI_paper, MARTINI_jos} is widely used. Designed for high-resolution smoothed particle hydrodynamics (SPH) simulations, it has been widely used to compare the properties of well-resolved \hi\ observations and simulated \hi\ distributions \citep[e.g.][]{ASymba,Bilimogga2022}. At low particle numbers, however, the effects of shot noise begin to become apparent in the mocks, notably on along the spectral axis.

Also required for comparisons between simulations and observations is a method to quantify galaxy morphology. One approach is the Concentration-Asymmetry-Clumpiness ``CAS"  \citep{Conselice2003} set of parameters. These parameters have been shown to be reliable for relatively high-resolution \hi\ observations \citep{Holwerda2011,Giese2016,Bilimogga2022}. Notably, spatial \hi\ asymmetries in galaxies are connected to interactions \citep{Holwerda2011}, accretion events \citep{Sancisi2008} and starbursts \citep{Lelli2014} for various samples of gas-rich galaxies. On the simulations side, \citet{Gensior2024} found that \hi\ spatial asymmetry is a promising observable to compare different simulation models with observations.

Despite their utility at high resolution, both \hi\ and optical spatial asymmetries wane at low angular resolutions, and noise introduces important systematic effects \citep{Giese2016, Thorp2021}. For example, using mock galaxies in \hi\ from the EAGLE simulation, \citet{Bilimogga2022} found that both spatial and spectral asymmetries are robust only at high resolution and signal-to-noise. In particular, the resolution requirements for robust spatial asymmetries strongly limit their application to \hi\ maps of galaxies from untargeted surveys such as the Widefield ASKAP L-band Legacy All-sky Blind surveY (WALLABY; \citealt{WALLABY0, WALLABY2022, WALLABY2022B}), which will image an unprecedented number of detections, most of which will however be only marginally spatially resolved. 

By contrast, spectral asymmetries such as lopsidedness \citep{Haynes_1998, Espada2011,Bok2019, Watts2023} and channel-by-channel profile asymmetries \citep{Deg2020, Reynolds2020} are useful for characterizing asymmetries with low spatial resolution but high spectral resolution that is characteristic of deep \hi\ surveys such as Deep Investigation of Neutral Gas Origins (DINGO; \citealt{dingo}) and Looking At the Distant
Universe with the MeerKAT Array (LADUMA; \citealt{laduma}).  

To explore asymmetries as a tool for upcoming \hi\ surveys, \citet{ASymba} introduce ASymba: Asymmetries of \hi\ in \textsc{Simba} Galaxies, which uses the SIMBA cosmological simulation suite \citep{SIMBA} to connect measured asymmetries to the physical drivers of galaxy evolution. \citet{ASymba} find that spectral asymmetries of SIMBA mock cubes correlate strongly with \hi\ mass, as well as with the number of mergers that a galaxy has undergone.
However, the high-resolution merging galaxy simulations by \citet{Deg2020} show that spectral asymmetries depend on observing parameters such as galaxy sky orientation and inclination, muddling their connection to the underlying galaxy structure. Using an alternative spectral \hi\ asymmetry metric, \citet{Watts2020b} compared galaxies from the TNG simulation \citep{TNG} to show that, while in many cases there is a connection with environment \citep[satellites are more asymmetric than centrals as a population, as seen in observations; e.g.,][]{Watts2020b}, central galaxies can also be asymmetric.

In light of the shortcomings of spatial and spectral asymmetries applied to \hi\ surveys, \citet{A3D} developed a 3D asymmetry measure that leverages the full dimensionality of the $\hi$ datacubes, outperforming spatial and spectral asymmetries for marginally spatially resolved detections in surveys such as WALLABY. \citet{A3D} also apply a squared differences background correction approach that provides a significant improvement to the calculation of asymmetries in the low signal-to-noise ($S/N$) regime compared to more conventional absolute differences (see also \citealt{rmsasym} and \citealt{Wilkinson2024}).

While there are many works that examine morphometrics in simulations (e.g \citealt{Lotz2008,Lotz2010,Abruzzo2018,ASymba,Bilimogga2022,Thorp2021,Wilkinson2024}) and in observations (e.g. \citealt{Conselice2003,Holwerda2011,Giese2016,Bok2019,Reynolds2020,Holwerda2023}), relatively few have directly compared asymmetries between them \citep[e.g.][]{CASCOMPARE, Deg2020}. With WALLABY pilot survey observations, a promising new 3D asymmetry technique, and an initial exploration of asymmetries in SIMBA now in hand, we proceed to develop a new method to create mock datacubes and compare measured WALLABY asymmetries to simulated SIMBA ones for the first time.

In this second paper of the ASymba series, we perform a quantitative comparison between 1D and 3D \hi\ asymmetries measured for spatially-resolved WALLABY Pilot Survey detections and WALLABY-like mock cubes derived for simulated SIMBA galaxies.  

Section \ref{sec:Datasets} describes the WALLABY and SIMBA datasets that we analyze.  Section \ref{sec:MockCubes} describes our process for generating mock cubes using a Scanline Tracing approach, and its impact on asymmetries measured from noiseless WALLABY-like cubelets. Next Section \ref{sec:WALLABYMock}, presents our mock WALLABY sample from SIMBA and compares it to the WALLABY sample. Finally, Section \ref{sec:conclusion} discusses our results.

\section{Datasets}\label{sec:Datasets}
\subsection{WALLABY}

WALLABY is an untargetted $\hi$ survey on the Australian Square Kilometre Array Pathfinder (ASKAP; \citealt{Hotan2021}) covering about 14000 square degrees of the southern sky.  It has a spatial resolution of $30\arcsec$, a spectral resolution of $18.5~\rm{kHz}$ (which corresponds to $\sim 4~\kms$ at $z \sim 0$), and a target sensitivity of $1.6~\rm{mJy~beam}^{-1}$, allowing it to detect the $\hi$ content of $\sim2\times10^{5}$ galaxies \citep{WALLABY2022,WallabyPDRPhase2}.  Simulated predictions from \citet{WALLABY0}, adjusted for the survey area in \citet{WALLABY2022} and \citet{WallabyPDRPhase2} suggest that $\sim2300$ galaxies will be spatially resolved by more than 5 beams, with $>10^{4}$ being marginally resolved.

WALLABY Pilot observations of a number of fields have been released in \citet{WALLABY2022} and \citet{WallabyPDRPhase2}, consisting of $\sim2400$ detections and $236$ kinematic models.  A WALLABY detection may consist of more than a single galaxy due to limitations in the resolution, $S/N$ and source finding. At only 1\% of the total WALLABY volume, the pilot observations already comprise the largest sample of uniformly analyzed interferometric observations of the $\hi$ content in galaxies.  This is an ideal sample for studying with morphometrics \citep{Holwerda2023}, which is why we have chosen it as our testbed for comparing simulated asymmetries with observations.

\subsection{SIMBA}
\label{sec:SIMBA}
\label{ssec:selection}

SIMBA \citep{SIMBA} is a cosmological simulation that runs on the \textsc{GIZMO} hydro-dynamical solver \citep{GIZMO} in Meshless Finite Mass (MFM) mode.  The simulation uses the \textsc{GRACKLE} library \citep{Grackle} to implement radiative cooling and photoionization heating. The $\hi$ fraction of the gas is evolved with the simulation instead of computed in post-processing, following the method of \citet{Dave2017}.  We have chosen to utilize the $50\ \mathrm{Mpc}$ box for this work, which contains $512^{3}$ particles in a comoving $\left(50\ h^{-1}~\mathrm{Mpc}\right)^3$ volume with a gas particle mass resolution of $1.82\times10^7\ \mathrm{M}_{\odot}$. In this box, the minimum full width at half-maximum (FWHM) smoothing length of the particles is $\sim 0.7~\mathrm{kpc}$.

The choice of the $50~\mathrm{Mpc}$ box is motivated by the need for a large number of well resolved \hi\ galaxies in the mass range of the WALLABY detections (see Sec. \ref{ssec:wallaby_asym}).  The other iterations of the $50\ \mathrm{Mpc}$ simulation with different feedback methods (for example with no AGN), would also allow future studies of how those physics drive changes in morphology.

Galaxies are identified using the CAESAR catalog. To avoid effects from poor numerical resolution, we follow the selection criteria of \citet{GlowackiSelection} to select systems for this study, namely:
\begin{itemize}
    \item $M_{\hi}>10^9~M_{\odot}$,
    \item $M_{\star} > 5.8\times10^8~M_{\odot}$,
    \item $\mathrm{sSFR}>1\times10^{-11}~\mathrm{yr}^{-1}$,
\end{itemize}
and to ensure that only galaxies that are numerically well resolved are selected, we impose:
\begin{itemize}
    \item $N_{2\mathrm{kpc}}>100$,
\end{itemize}
where $N_{2\mathrm{kpc}}$ is the number of fluid elements with smoothing length FWHM below $2\,\mathrm{kpc}$. 789/5218 SIMBA galaxies fall within these criteria.

\section{SIMBA Mock Cubes}\label{sec:MockCubes}

To compare WALLABY detections to SIMBA systems, it is necessary to generate mock observations of the latter that have similar noise and resolution to that of the WALLABY datacubes.

However, we show in Sec. \ref{ssec:scanline} that widely used methods of generating mock \hi\ cubes tend to produce significant artificial shot noise contributions to noiseless WALLABY-like mocks. As a result, we present a new Scanline Tracing method for generating mock cubes from simulations in Sec. \ref{sec:RTC}. Section \ref{ssec:SIMBA_A} illustrates the impact of adopting this new method on the asymmetry measures defined by \citet{A3D} for a suite of noiseless mock SIMBA detections.

\subsection{Mock Observations}
\label{ssec:SIMBA}
Cosmological simulations evolve sets of discrete elements that represent continuous fluids such as gas. The smoothed particle hydrodynamics \citep[SPH, see][]{SPHLucy, SPHGingoldMonaghan} and MFM techniques of interest for this paper are two methods of discretizing the fluid equations. In our case, a mock \hi\ observation consists in the transformation of the numerical elements of a simulation into an \hi\ cube measured as $f_{i,j,k}$ where the $i,j,k$ indices denote spatial and spectral indices allowing to specify the voxel of the datacube. While the cube is measured in discrete coordinates, it will be easier in the following  derivations to write its coordinates in terms of continuous sky variables $f(x,y;v)$. 

To model gases, such as \hi, SPH methods use a kernel $W$ to interpolate a given field $\phi$ between particles:
\begin{equation}
\label{eq:SPH}
    \phi_s\!\left(\bm{x}\right)= \sum_n \phi_n V_n W\!\left(\bm{x}\!-\!\bm{x}_n;h_n\right),
\end{equation}
with $\phi_s$ being the smoothed field, $\bm{x}$ being the position, $V$ the volume, $h$ the smoothing length, and the $n$ indices iterate over the particles and their properties. At sufficient resolution, $\phi_s\approx\phi$. 
The MFM technique introduces a local normalization to the total kernel; the $n^{\mathrm{th}}$ particle's MFM kernel $\psi_i$ relates to the SPH kernel $W$ according to the definition
\begin{equation}
    \psi_n\!\left(\bm{x}\right) \equiv \dfrac{ W\!\left(\bm{x}\!-\!\bm{x}_n;h_n\right)}{\sum_m W\!\left(\bm{x}\!-\!\bm{x}_m;h_m\right)},
\end{equation}
where $m$ iterates over all particles. MFM smoothed fields are calculated akin to SPH:
\begin{equation}
\label{eq:MFM}
    \phi_s\!\left(\bm{x}\right) = \sum_n \phi_n \psi_n\!\left(\bm{x}\right).
\end{equation}

\subsubsection{Particle-Built Mock Cubes}
\label{ssec:scanline}

\textsc{MARTINI} \citep{MARTINI_soft, MARTINI_paper, MARTINI_jos} is a widely used method for generating mock \hi\ datacubes.  By default, MARTINI extends Equation~\ref{eq:SPH} to an assumed particle Gaussian distribution along the frequency axis; this distribution is evaluated  individually and then added to produce the particle method's intensity $f_p$ at a given spatial and spectral coordinate:
\begin{equation}
    f_p\!\left(x,y;v\right) \equiv \sum_n s_n\!\left(v\!-\!v_n;\sigma_n\right) V_n \int_{\mathbb{R}} W\!\left(\bm{x}\!-\!\bm{x}_n;h_n\right)\mathrm{d}z,\label{eq:MARTINI}
\end{equation}
with $\sigma_n$ denoting the broadening corresponding to the particle temperature, $s_n$ being a Gaussian spectrum corresponding to the particle and $z$ being the line of sight axis.

\begin{figure*}[htbp]
    \centering
    \includegraphics{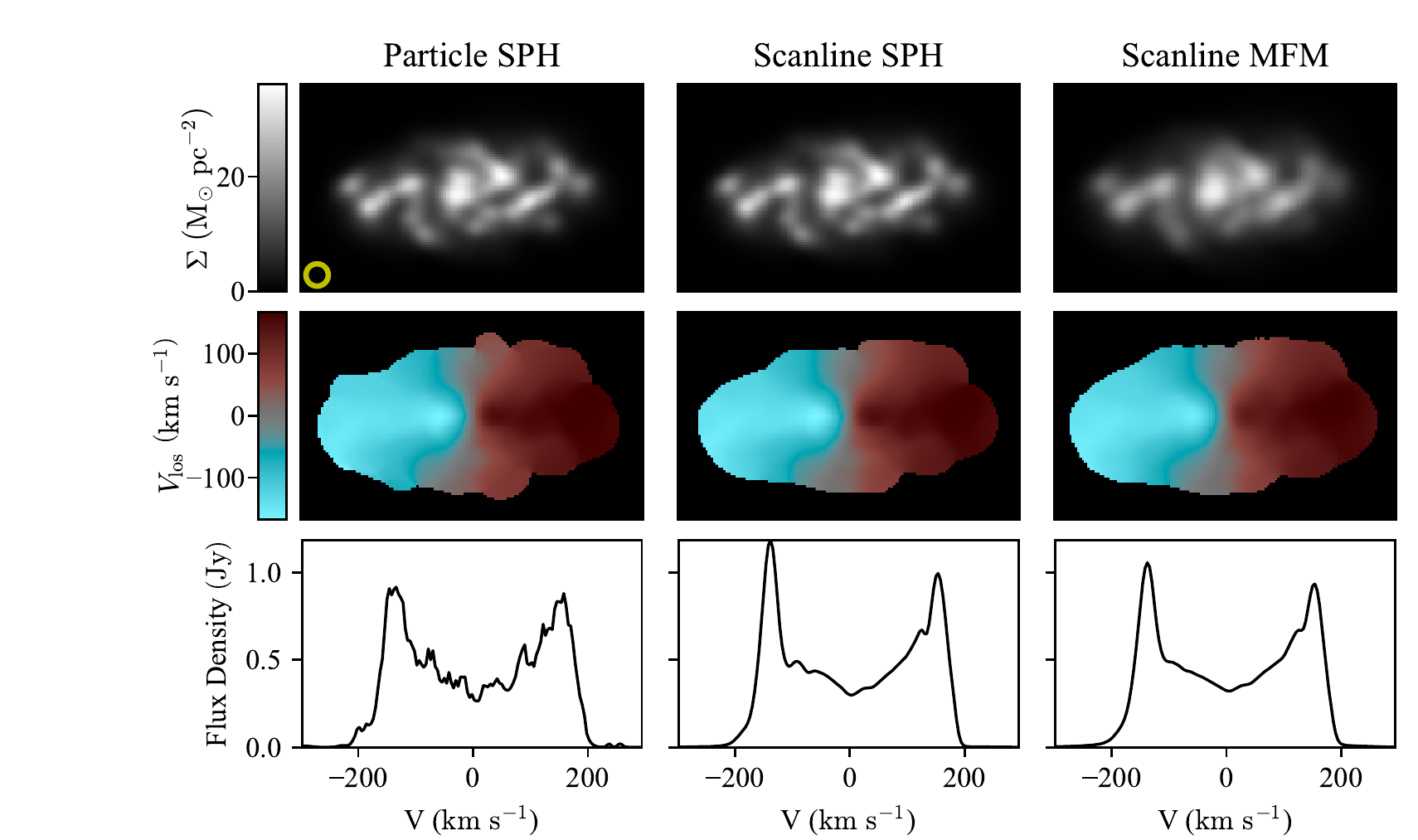}
    \caption{Moment 0 maps (top), moment 1 maps (middle) and spectra (bottom) for the Particle SPH (left) and Scanline Tracing SPH (middle) and MFM (right) constructed from noiseless mock $\hi$ cubes of Galaxy 467 of the SIMBA $50\ \mathrm{Mpc}$ simulation taken at a distance of $20$ Mpc with WALLABY observation parameters. The ring in the bottom left corner of the first moment 0 map shows the beam FWHM. For the moment 1 maps, the cube is masked at a threshold of $1.6\ \mathrm{mJy}\ \mathrm{beam}^{-1}$ (approximately the WALLABY noise RMS) and we use the CosmosCanvas \citep{CosmosCanvas} color map. The galaxy is composed of $\sim 1600$ collisional particles and has an $\hi$ mass of $\sim 1.8\times10^{10}\ \mathrm{M}_{\odot}$.}
    \label{fig:gal467}
\end{figure*}

\begin{figure*}[htbp]
    \centering
    \includegraphics{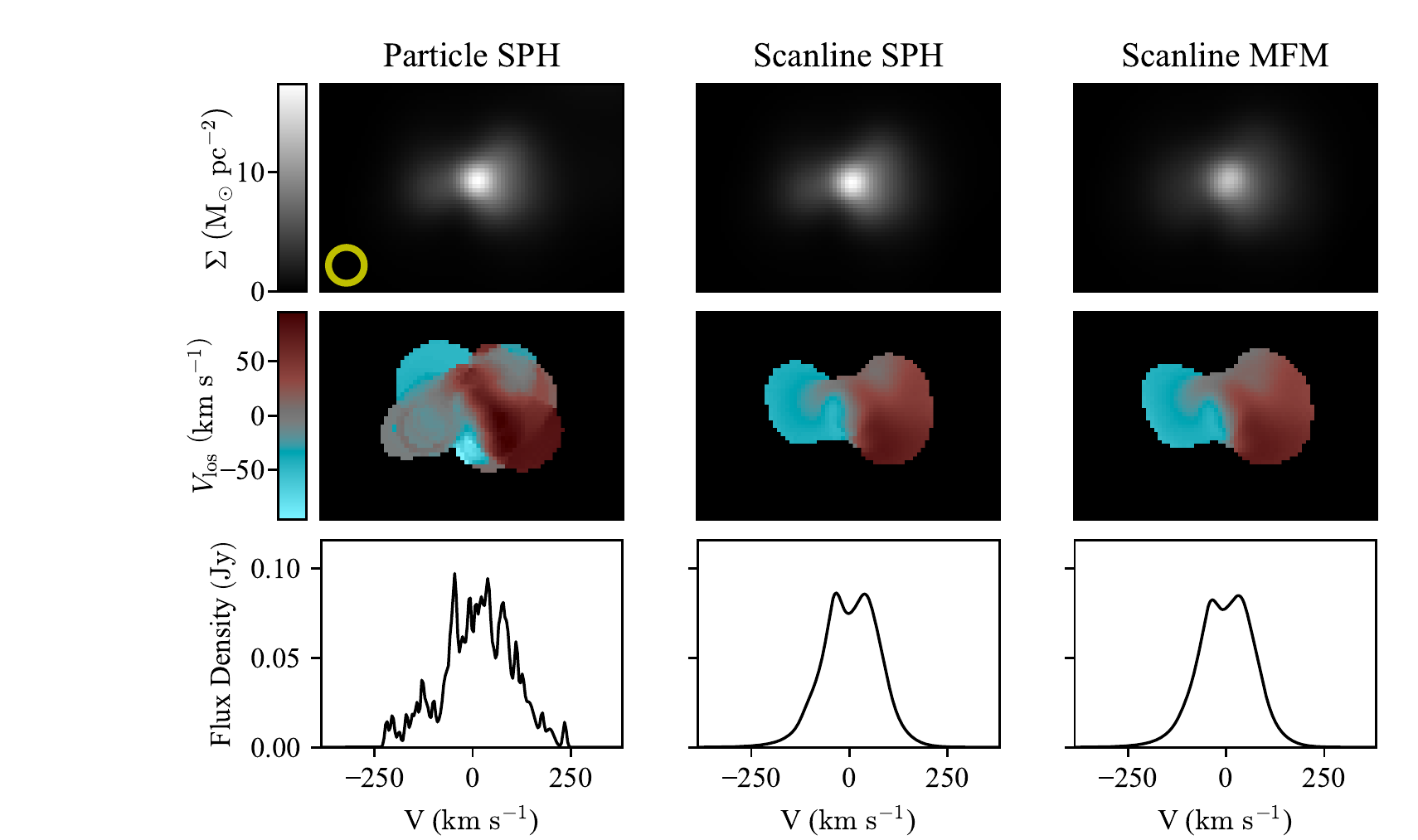}
    \caption{Same as Figure~\ref{fig:gal467}, but for a noiseless, WALLABY-like mock of Galaxy 1531. The galaxy is composed of $\sim 1100$ collisional particles and has an $\hi$ mass of $\sim 1.5\times10^{9}\ \mathrm{M}_{\odot}$.}
    \label{fig:gal1531}
\end{figure*}

\begin{figure}[htbp]
    \centering
    \includegraphics{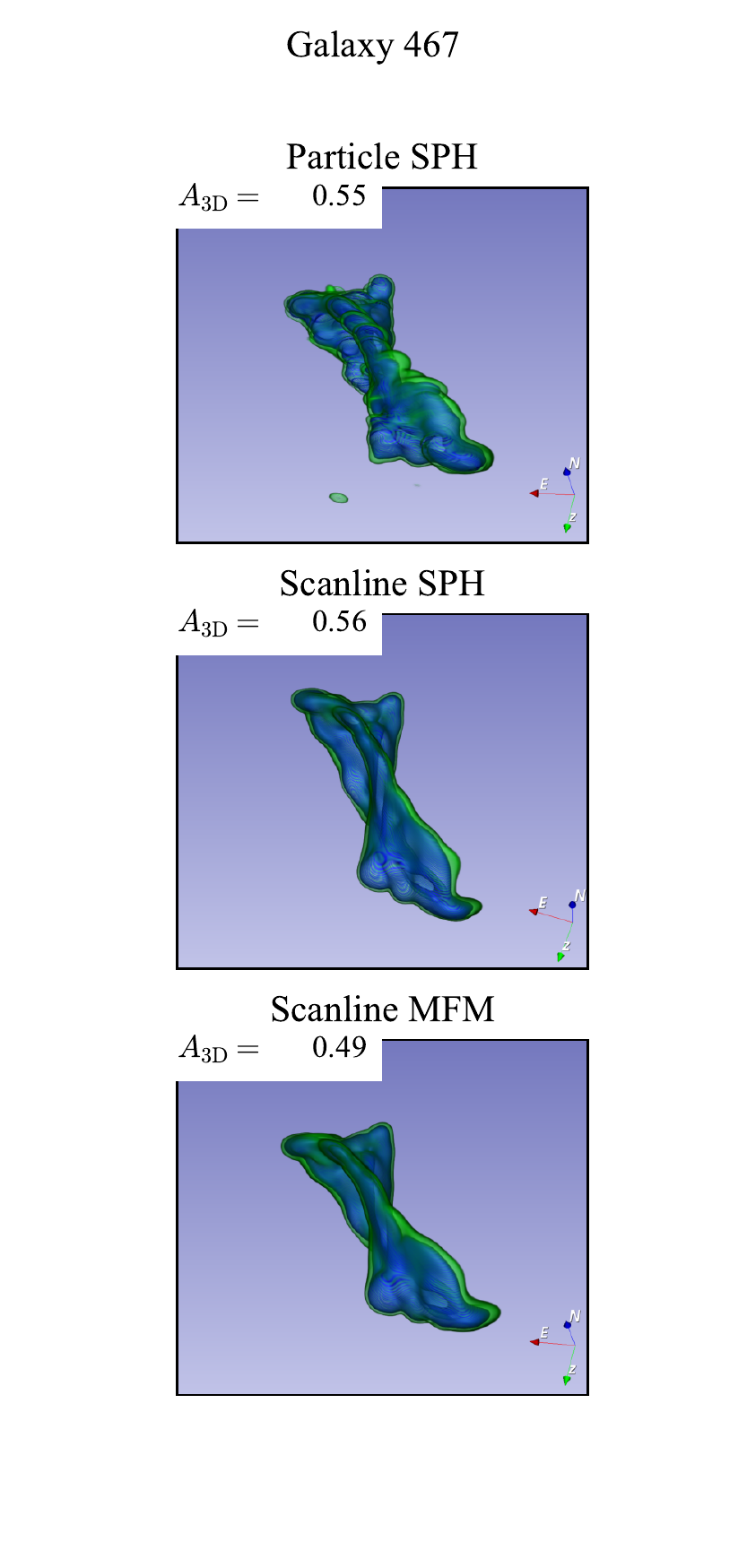}
    \caption{\textsc{SlicerAstro} view of noiseless mocks of the $\hi$ cube of Galaxy 467 presented in Figure \ref{fig:gal467} for the particle SPH (top), scanline SPH (middle), and scanline MFM (bottom). The E-N-Z axes show the RA-DEC-Velocity axes respectively, with +Z indicating the approaching velocity. The values of $A_{3\mathrm{D}}$ for each mock, computed about the minimum of the potential, are given in the top-left corner of each panel.} 
    \label{fig:slicer467}
\end{figure}

\begin{figure}[htbp]
    \centering
    \includegraphics{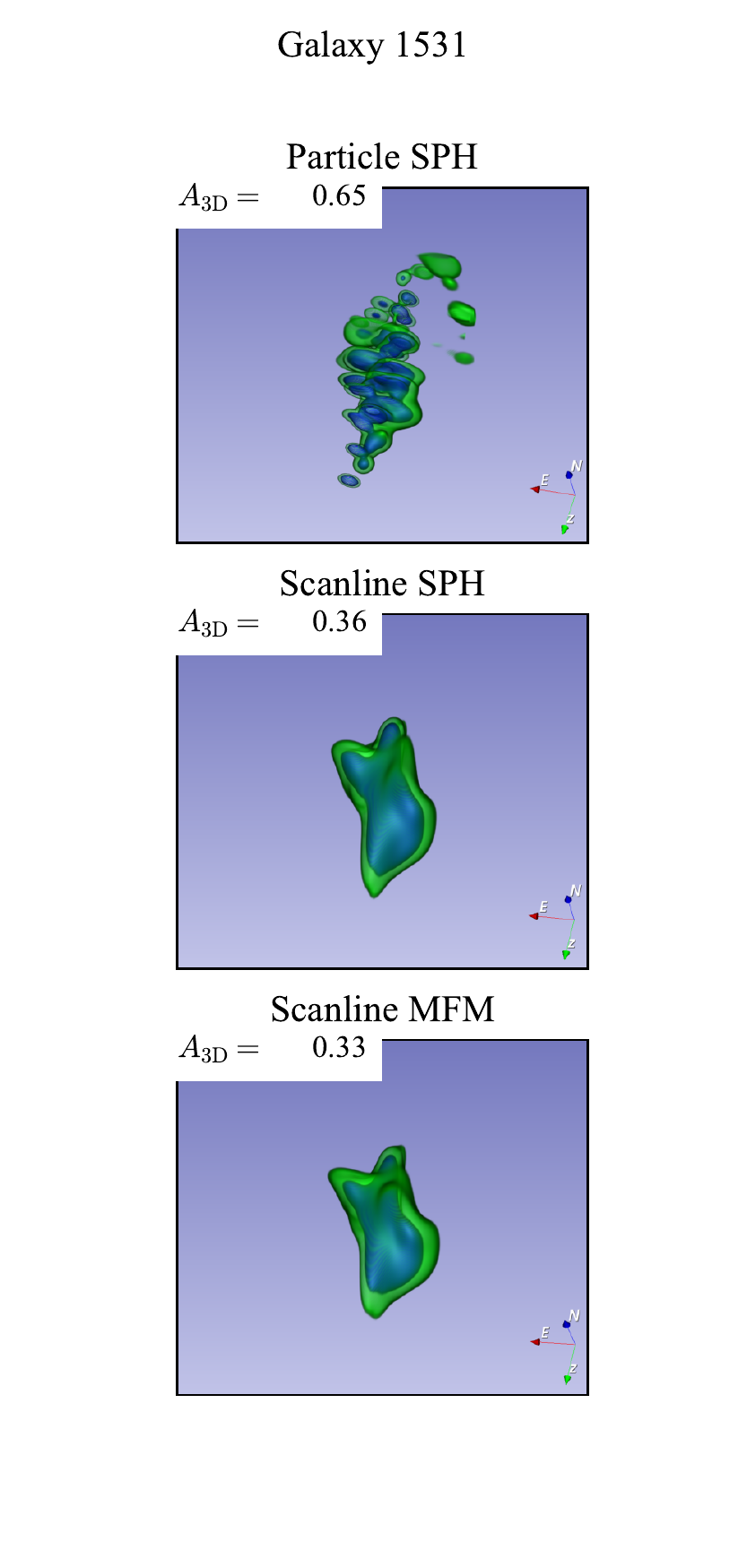}
    \caption{Same as in Figure~\ref{fig:slicer467}, but for  Galaxy 153 presented in Figure \ref{fig:gal1531}.} 
    \label{fig:slicer1531}
\end{figure}

While this method has been used to study asymmetries in both the EAGLE simulation \citep{Bilimogga2022}, and the SIMBA simulation \citep{ASymba}, the approximations in Equation~\ref{eq:MARTINI} can cause issues.  To illustrate, we have selected two representative SIMBA sources, Galaxy\~467 and Galaxy\~1531, and generated noiseless mock observations of them using \textsc{MARTINI}.  Moment 0 (intensity) and Moment 1 (velocity) maps and spectra for the two \textsc{MARTINI} realizations are shown in the left-hand columns of Figures \ref{fig:gal467} and \ref{fig:gal1531}, while the upper panels of Figures \ref{fig:slicer467} and \ref{fig:slicer1531} show 3D renderings of the \textsc{MARTINI} cubes generated through the use of \textsc{SlicerAstro} \citep{SlicerAstro}.  

Galaxy 467 has a relatively high particle number ($\sim1600$), but the spectrum in the lower left panel of Figure \ref{fig:gal467} is fairly jagged. We attribute this to shot noise caused by the superposition of gaussians. While SPH and MFM elements tend to be spatially distributed in a meaningful way, this is not necessarily the case in velocity space. Certain parts of the velocity field may be over or under sampled, leading to the observed jagged features. This noise is more significant at the lower particle number in Galaxy 1531 (lower left panel of Figure \ref{fig:gal1531}).  The jaggedness of these spectra will artificially increase both 1D and 3D asymmetries.  This issue is not present in the \citet{Bilimogga2022} study of EAGLE asymmetries due to both the higher particle resolution of EAGLE as well as their application of Hanning smoothing to their spectra.

The spectral jaggedness of these \textsc{MARTINI} cubes can be understood by examining the behavior of a pair of particles separated in velocity space. Applying Equation \ref{eq:MARTINI} to these particles leads to distinct peaks when the particle velocities are sufficiently separated relative to their temperature, regardless of their smoothing lengths. If the velocity, density and temperature fields were instead evaluated separately according to either, for SPH, Equation \ref{eq:SPH} or, for MFM, Equation \ref{eq:MFM}, the spectrum would be washed out of spectral features for sufficiently large smoothing lengths, regardless of the particle temperatures and velocities. This is a consequence of the SPH and MFM interpolation schemes not being invariant under transformation. For example, if a field $\phi$ were interpolated and then squared, it would yield a different field than if its square were directly interpolated:

\[\left(\sum \phi_n V_n W\!\left(\bm{x}\!-\!\bm{x}_n;h_n\right)\right)^2 \neq
    \sum \phi_n^2 V_n W\!\left(\bm{x}\!-\!\bm{x}_n;h_n\right).\]

Effectively, Equation \ref{eq:MARTINI} introduces shot noise instead of the expected oversmoothing from SPH at low resolutions.  According to Equation \ref{eq:SPH} there should be intermediate regions of the gas where the maximal spectral intensity corresponds to mediating velocities. However, this behavior is not enforced by Equation \ref{eq:MARTINI} unless the particles sample these regions of the velocity field. Additionally, Equation \ref{eq:MARTINI} is unlikely to create a spectrum that is locally Gaussian, whereas Equation \ref{eq:SPH} enforces this. The appearance of the different points in the spectrum can be attributed to the spectral separation of the particles being lower than their dispersion $\sigma^2_n$. While the particle-based spectra can be smoothed by artificially injecting additional velocity dispersion into each particle, this is unphysical.  Instead, we develop a different method of computing mock \hi\ datacubes that directly sample the fields.

\subsubsection{Scanline Tracing}
\label{sec:RTC}

The \textsc{MARTINI} realizations of Galaxies 467 and 1531 shown in Figures \ref{fig:gal467} - \ref{fig:slicer1531} and discussed in Sec. \ref{ssec:scanline} suggest that for our specific asymmetry analysis of SIMBA galaxies, a new method of generating mock \hi\ cubes is needed.  To that end we adopt a \textit{scanline} method, where the underlying fields are sampled via scanlines cast through simulation space along `observed' lines of sight. For our purposes, the small angle approximation holds and the scanlines are held parallel. To simplify the math, we will hold the line of sight as being in the $z$ axis.  The scanlines sample the fields at set intervals, resulting in a 3D grid of the simulation space that can then be collapsed into a datacube.

The spectrum is calculated as a function of the fields for a given position:
\begin{equation}
\label{eq:localspectrum}
    s\left(\bm{x},v\right)\equiv
    \dfrac{\rho\left(\bm{x}\right)}{\sqrt{2\pi\sigma^2\!\left(\bm{x}\right)}}\exp\left(-\frac{\left(v-v_z\!\left(\bm{x}\right)\right)^2}{2\sigma^2\!\left(\bm{x}\right)}\right),
\end{equation}
with $\rho\left(\bm{x}\right)$, $v_z\left(\bm{x}\right)$, $\sigma\left(\bm{x}\right)$ being the $\hi$ density, line of sight velocity and thermal velocity dispersion fields of the fluid at a given position $\bm{x}$ and $v$ being the velocity at which the spectrum's intensity is measured. 
In the case of MFM, the velocity is treated such that the quantity of movement of a fluid element is defined as equal to that of the integral of the quantity of movement of the fluid over the particle's effective volume, but we consider the interpolation of the velocity field to also be reasonable.

The spectra encountered by the scanlines at each step $\mathrm{d}z$ along the line of sight are collapsed into the spectrum at a given coordinate $x,y$ on the sky. For a pixel of width $\Delta x$ and height $\Delta y$, this gives
\[
    f\left(x,y;v\right)
    =
    \int_{x-\frac{\Delta x}{2}}^{x + \frac{\Delta x}{2}}\!
    \int_{y-\frac{\Delta y}{2}}^{y + \frac{\Delta y}{2}}\!
    \int_{\mathbb{R}}
    s\!\left(\bm{x}',v\right)
    \mathrm{d}x'\mathrm{d}y'\mathrm{d}z'.
\]
To approximate this integral, we evaluate the local smoothed spectrum $s_s\left(\bm{x},v\right)$ at set intervals with sampled $\rho_s\left(\bm{x}\right), \sigma^2_s\left(\bm{x}\right), v_s\left(\bm{x}\right)$ fields using Equation~\ref{eq:SPH} or \ref{eq:MFM} as appropriate. 

The density can be measured according to the SPH formalism:
\begin{equation}
    \rho_s\left(\bm{x}\right) = \sum_n {m}_n W\left(\bm{x}-\bm{x}_n;h_n\right),
\end{equation}
with $m_n$ being the mass of the $n$th particle. Since only the $\hi$ mass is of interest to us, we calculate the densities using the particle $\hi$ masses instead of their total masses.
This equation ensures that the mass is conserved for a kernel $W$ normalized under integration. This version of the density calculation agrees with the particle-based densities.

In MFM, the conserved quantities are considered as averaged over the particle volume. The particle density can't be measured directly because the kernel is not normalized under integration. The particle effective volume is defined as
    \[V_n\equiv\iiint_{\mathbb{R}^3} \psi_n\left(\bm{x}\right) \mathrm{d}^3\bm{x},\]
such that the particle density is $\rho_n={m}_n/V_n$. However, this would require the evaluation of the integral of the kernel, which is non-elementary. For particles fully contained by the Scanline Tracing box of domain $\mathbb{B}$, the volume can be estimated by summation. For particles that are not contained by the box, the volume is scaled up according to the volume across which the particle's kernel is non-zero. The volume is then evaluated:
\begin{equation}
\label{eq:MFMVolume}
    V_n \approx
    \begin{cases}
       \sum_{\psi_n\neq0} \psi_n \Delta V \ \ \text{if}\ \bm{x}_n+\bm{h}_n \in \mathbb{B}\ \forall\ \bm{h}_n \\
       \\
       \dfrac{ \iiint_{W\left(\bm{x};h_n\right) \neq 0}  \mathrm{d}^3\bm{x} \sum_{\psi_n\neq0} \psi_n \Delta V}{\sum_{\psi_n\neq0} \Delta V}
     \end{cases},
\end{equation}
where $\Delta V$ is the volume element of the 3D grid across which the fields are evaluated, $\bm{h}_n$ is an arbitrary vector of length $h_n$. The integral corresponds to the volume where the kernel is non-zero. Using for example a kernel where $W(\bm{x}-\bm{x}_n;h_n)=0\ \forall\ \left|\bm{x}-\bm{x}_n\right|>h_n$, the integral gives the sphere $4\pi h_n^3/3$.
With the volume, the density can be calculated locally with
\begin{equation}
\label{eq:MFMDensity}
    \rho_s\left(\bm{x}\right) = \sum_n \dfrac{m_n}{V_n} \psi_n \left(x\right).
\end{equation}
When particles fall completely inside the sampled box, Equation \ref{eq:MFMVolume} ensures that Equation \ref{eq:MFMDensity} will always give a density contribution consistent with the particle total mass if the density is evaluated at the same points as those used to evaluate the volume. The other fields are interpolated using Equation \ref{eq:MFM}.

The SIMBA simulation uses the cubic spline kernel:
\begin{subequations}
\begin{equation}
        W\left(\bm{x};h\right)=\frac{1}{h^3}w\left(\frac{\left|\bm{x}\right|}{h}\right),
\end{equation}
\begin{equation}
        w(q) \equiv
        \dfrac{8}{\pi}
    \begin{cases}
       1-6q^2\left(1-q\right) &\quad\text{if } q\le1/2\\
       2\left(1-q^3\right)^3 &\quad\text{if } 1/2<q\le1\\
       0
     \end{cases}
     .
\end{equation}
\end{subequations}
Once the fields are obtained, Equation \ref{eq:localspectrum} is evaluated at each spectral channel $\Delta v$. The sampled spectra are then added to approximate the integral along the scanline:
\begin{equation}
    f\!\left(x,y;v\right)
    \approx
    \Delta\! V\!
    \sum_{i}^{}
    \!
    \frac{\rho_s\!\left(x,y,i\Delta\! z\right)}{\sqrt{2\pi\sigma^2_s\!\left(x,y,i\Delta\! z\right)}}e^{-\frac{\left(v-v_s\!\left(x,y,i\Delta\! z\right)\right)^2}{2\sigma^2_s\!\left(x,y,i\Delta\! z\right)}},    
\end{equation}
with the summation taken over the number of times each scanline samples the fluid fields. This allows each pixel to have spectral structure, but enforces that the spectrum of the gas is locally Gaussian. Finally, the pixels are assembled from the rays and are convolved according to a desired beam to create a datacube. The pixels are often over-sampled to ensure that their size doesn't alter the spectrum. 

For our specific case, we construct cubes at resolutions of half the minimum smoothing length. After being calculated at this resolution, cubes are downsampled to the desired resolution using the \textsc{Open CV2} pixel-area relation INTER\_AREA \citep{opencv_cite}.  The cube is constructed for a region encompassing all the gas that can contribute to a moment 0 map above the noise level at a 10 Mpc distance extended by two beams at the specified observation distance. Relaxing these parameters does not meaningfully alter the results of Sections \ref{ssec:SIMBA_A} and \ref{sec:WALLABYMock}.

With this formalism, it is possible to construct mock cubes using Scanline Tracing for either SPH or MFM.  The middle columns of Figures \ref{fig:gal467} and \ref{fig:gal1531} show the moment maps and spectra for noiseless Scanline SPH realizations of Galaxies 467 and 1531 from SIMBA, while the right-hand column shows the Scanline MFM realization.  Similarly, the middle and lower panels of Figures \ref{fig:slicer467} and \ref{fig:slicer1531} show the 3D renderings of the SPH and MFM realizations.  The Scanline Tracing mocks for both Galaxy 467 and 1531 show little evidence of shot noise.  For Galaxy 1531, the Scanline mocks have a narrow, double-peaked profile.  Moreover, in this lower particle number, lower mass regime, the Scanline Tracing method dramatically changes the moment 1 maps.  This demonstrates that the Scanline tracing causes this change and not the assumption of using SPH or MFM to construct the cubes.

The comparison of the Scanline MFM realizations to the Scanline SPH realizations has a much smaller effect on the mock cube.  The spectra are slightly smoother and the moment 0 maps appear to be slightly less well resolved.  This is due to the MFM kernel normalization distributing more gas to regions where there is less particle overlap. 

\subsection{SIMBA Asymmetries}
\label{ssec:SIMBA_A}

To fully understand the differences between Particle generated cubes and Scanline Tracing generated cubes for measuring asymmetries, we made mock observations of the 789 SIMBA galaxies that satisfy our criterion in Sec. \ref{ssec:selection} using both the particle-based \textsc{MARTINI} and our new Scanline Tracing MFM method. The mock cubes have a resolution of $30\arcsec$ and $4~\kms$, matching WALLABY, though they are noiseless to better probe the impact of the different approaches. For this test, all galaxies are placed at a distance of 20 Mpc (where 1 kpc = $10\arcsec$ on the sky), with an inclination of $60^{\circ}$ (calculated based on the angular momentum vector).    This distance allows for sufficient resolution elements for calculating the spatial asymmetry.  

With the two suites of noiseless SIMBA cubes generated, we calculate their spectral (1D), spatial (2D), and 3D asymmetry using the 3D Asymmetries in data CubeS (\textsc{3DACS}; \citealt{A3D}) code.  The asymmetry formula in the idealized noiseless case is
\begin{equation}\label{Eq:Asym}
    A^2= \frac{P}{Q} \,\,\,,
\end{equation}
where $P$ and $Q$ are the squared odd and even parts of the chosen distribution respectively:
\begin{subequations}
\label{eq:AsymParts}
    \begin{equation}
        P=\sum_{\bm{i}}\left(f_{\bm{i}}-f_{\bm{i}}\right)^2\ \,\,\,,
    \end{equation}
    \begin{equation}
        Q=\sum_{\bm{i}}\left(f_{\bm{i}}+f_{\bm{i}}\right)^2.
    \end{equation}
\end{subequations}
The bold $\bm{i}$ indices denote that we can compute the asymmetry over any number of dimensions, collapsing the others beforehand, then computing the asymmetry using Equation \ref{eq:AsymParts} and \ref{Eq:Asym}:
\begin{subequations}
\label{eq:AsymDims}
    \begin{equation}
        f_{\bm{i}=(k)}=\sum_{i,j}f_{i,j,k}\ \ \mathrm{for}\ A_{1\mathrm{D}},
    \end{equation}
    \begin{equation}
        f_{\bm{i}=(i,j)}=\sum_{k}f_{i,j,k}\ \ \mathrm{for}\ A_{2\mathrm{D}},
    \end{equation}
        \begin{equation}
        f_{\bm{i}=(i,j,k)}\ \ \mathrm{for}\ A_{3\mathrm{D}}.
    \end{equation}
\end{subequations}
We note that $A_{1\mathrm{D}}$ as defined above is equivalent to the channel-by-channel asymmetry developed by \citet{Deg2020} and \citet{Reynolds2020}. 

A critical component of the asymmetry is the center point about which the pairs of fluxes are compared.  Both the \textsc{MARTINI} Particle mocks and our Scanline Tracing mocks are centered at the minimum of total potential spatially and the bulk velocity of the particles.  This center is computed according to the particle data, making it independent of the chosen mock method (SPH-Particle, MFM/SPH Scanline Tracing).  The \hi\ gas is not always centered at this potential minimum, which can cause large asymmetries.

Before proceeding to the full distribution of asymmetries, it is worth returning to our two example galaxies.  The upper left corner each panel in Figures \ref{fig:slicer467} and \ref{fig:slicer1531} provide the 3D asymmetry for the \textsc{MARTINI} Particle SPH, Scanline SPH, and Scanline MFM realizations of Galaxies 467 and 1531 respectively.  Given that $0 \leq A_{3\mathrm{D}} \leq 1$ in the noiseless case (see Section \ref{sec:WALLABYMock} for a discussion of the effect of noise), the differences between the Particle and Scanline MFM realizations are significant.  This is particularly true in the low particle number regime of Galaxy 1531.

\begin{figure}[htbp]
    \centering    \includegraphics{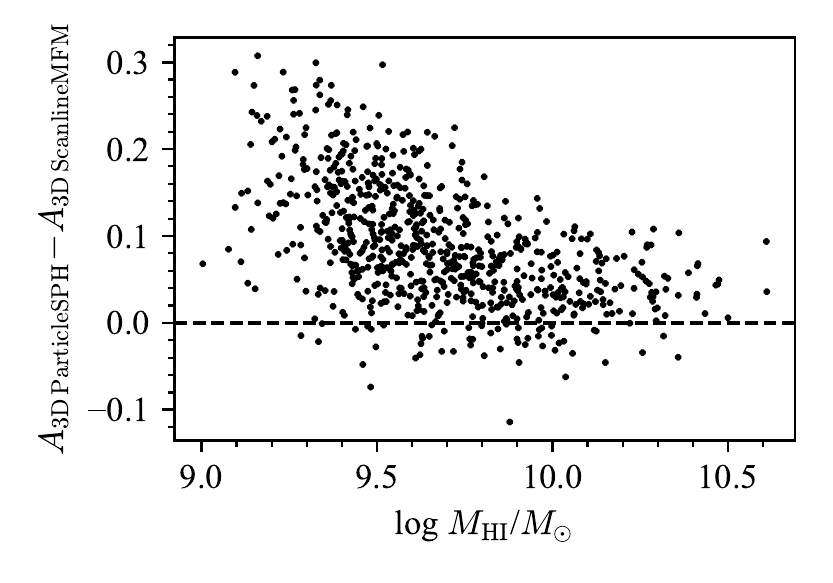}
    \caption{
    Plot of the difference of 3D asymmetry between the Particle SPH and MFM Scanline Tracing realisations of SIMBA mock cubes described in Section~\ref{ssec:selection} as a function of the cube \hi\ mass.}
    \label{fig:histVSA3D}
\end{figure}

This difference holds for the majority of SIMBA galaxies.  Figure~\ref{fig:histVSA3D} shows the difference between the Particle SPH 3D asymmetries and Scanline MFM 3D as a function of \hi\ mass.  The Particle SPH asymmetries are usually larger than their Scanline MFM counterparts. Particle SPH realizations are more asymmetric by as much as $\Delta A_{3\mathrm{D}} \sim 0.3$, mostly due to shot noise.  However, the differences decrease with increasing mass and therefore increasing particle number.  As such, at higher particle resolutions, the \textsc{MARTINI} Particle SPH realizations will likely converge to the Scanline MFM realizations, making the specific method of generating mock observations less critical. Since we wish to compare SIMBA galaxies with as few as 1000 particles to WALLABY detections, we exclusively use the Scanline Tracing MFM method of generating mock SIMBA cubes hereafter.

Before turning to WALLABY comparisons in Section~\ref{sec:WALLABYMock}, it is worth exploring the relationship between 1D, 2D, and 3D asymmetries for noiseless Scanline MFM mocks.  \citet{A3D} examined these measures using idealized mock cubes generated through a modified version of the Mock Cube Generator Suite (\textsc{MCGSuite}; \citealt{MCGcode}) code where asymmetries were introduced as first Fourier moments.
This type of asymmetry -- lopsidedness -- is discussed in \citet{Rix1995} and \citet{Zaritsky1997}. \citet{A3D} found that $A_{1\mathrm{D}}$, $A_{2\mathrm{D}}$, and $A_{3\mathrm{D}}$ increase linearly with lopsidedness, and thus that all asymmetries in their analysis are correlated. Here, we use SIMBA to explore whether or not these correlations persist for  asymmetries produced by simulated cosmological galaxy assembly.

\begin{figure}[htbp]
    \centering
    \includegraphics{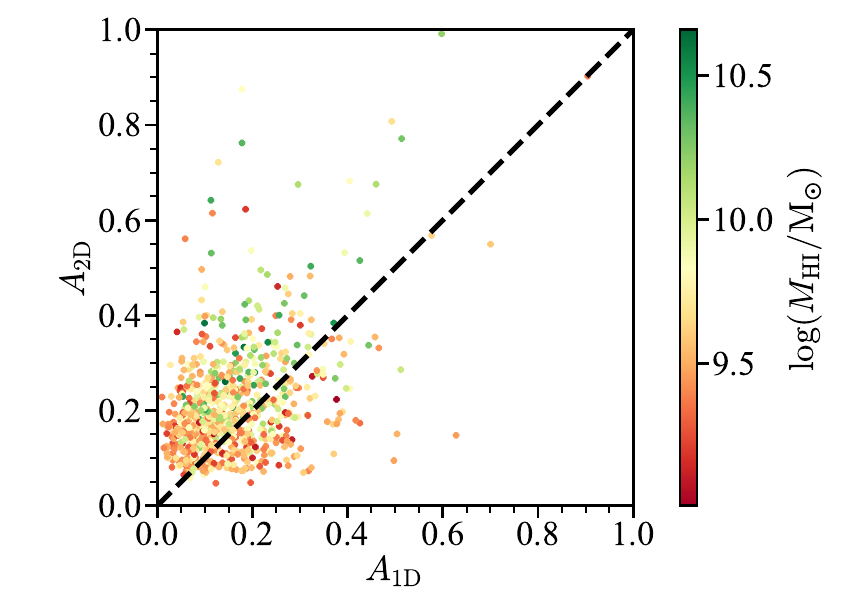}
    \caption{Comparison of 1D and 2D asymmetries of noiseless, WALLABY-like, Scanline-MFM mocks of SIMBA galaxies. The dashed black line shows the 1:1 relation. The points are coloured by $\hi$ mass, which correlates to resolution by virtue of the size-mass relation \citep{Wang2016}.}
    \label{fig:1D_vs_2D}
\end{figure}

In Figure \ref{fig:1D_vs_2D}, we plot the recovered 1D and 2D asymmetries for the Scanline Tracing MFM cubes, colored by their \hi\ mass. We find that spectral (1D) and spatial (2D) asymmetries are not strongly correlated, supporting the findings of \citet{Bilimogga2022}. It is possible that the drivers of spectral and spatial asymmetries are different, which would lead to uncorrelated asymmetries.  However, the measured asymmetry also depends on the adopted center point, and the \hi\ gas in SIMBA galaxies is not always centered at the minimum of the potential.  Secondly, asymmetry tends to be correlated with resolution, particularly in the low resolution regime \citep{Giese2016,A3D}.  The spatial and spectral resolutions are effectively independent.  At low spatial resolutions, the spatial asymmetry will go to zero, leaving the spectral asymmetry intact, which again will hide any possible correlations.  \citet{A3D} argue that robust measures of the 2D asymmetry require $\sim7$ or more resolution elements \citep[see also][]{Giese2016,Thorp2021}.  This requirement is satisfied for this particular sample of mock SIMBA galaxies at 20\~Mpc with a $30\arcsec$ beam, but it is not true for the majority of the WALLABY observations (see Sec. \ref{sec:WALLABYMock}).

\begin{figure}[htbp]
    \centering
    \includegraphics{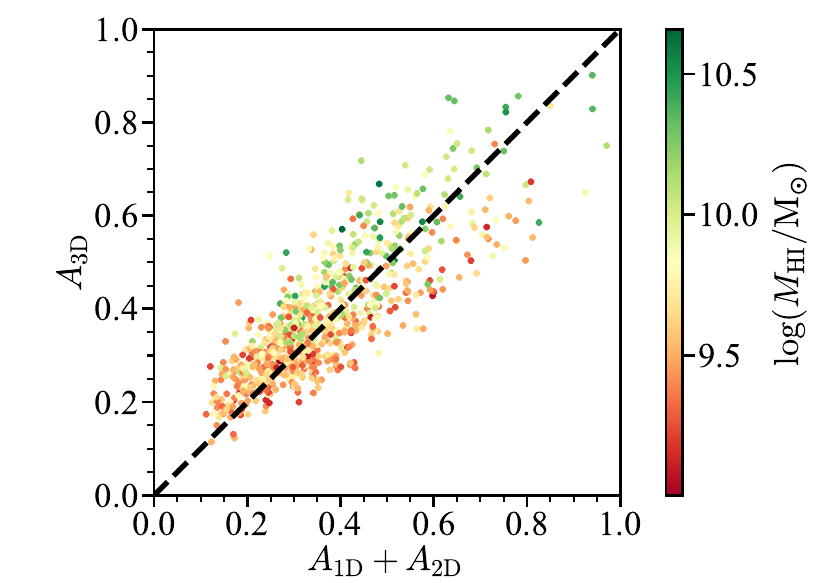}
    \caption{Same as in Figure~\ref{fig:1D_vs_2D}, except for the combination of 1D and 2D asymmetries plotted against 3D asymmetry. 10 \hi\ sources fall outside the $0,1$ range for the combined $A_{1\mathrm{D}}$+$A_{2\mathrm{D}}$. }
    \label{fig:A1D2D_A3D}
\end{figure}

The $A_{3\mathrm{D}}$ measure was designed to capture both spectral and spatial asymmetries in $\hi$ datacubes \citep{A3D}. This is illustrated in Figure \ref{fig:A1D2D_A3D}, which shows a comparison of $A_{3\mathrm{D}}$ to $A_{1\mathrm{D}}+A_{2\mathrm{D}}$. To focus on the most important range, we have cut the full dynamical range of $A_{1\mathrm{D}}+A_{2\mathrm{D}}$ which extends to 2, and 10 \hi\ mocks with $A_{1\mathrm{D}}+A_{2\mathrm{D}}>1$ are missing. Comparing the $A_{3\mathrm{D}}$ in this figure to either the 1D or 2D asymmetry distributions in Figure \ref{fig:1D_vs_2D} shows that the 3D asymmetries span a larger dynamic range than either 1D or 2D for the noiseless WALLABY-like SIMBA mocks.  We note that in Figure \ref{fig:1D_vs_2D}, the high and low mass mock cubes split along the 1:1 line, with the high mass cubes trending to higher $A_{3\mathrm{D}}$ for a given $A_{1\mathrm{D}}+A_{2\mathrm{D}}$ whereas the low mass mocks fall below this line. This bifurcation can be explained by resolution effects impacting the spatial (2D) component of the asymmetry, as, at fixed distances, $\hi$ mass is correlated to resolution through the $\hi$ size-mass relation \citep{Wang2016}. Figures \ref{fig:1D_vs_2D} and \ref{fig:A1D2D_A3D} demonstrate that the advantages of $A_{3\mathrm{D}}$ over $A_{1\mathrm{D}}$ and $A_{2\mathrm{D}}$ posited by \citep{A3D} using idealized mocks are also evident in more complicated cosmologically-generated mocks.

\section{Asymmetries and the Mock WALLABY Sample}
\label{sec:WALLABYMock}

We now quantitatively compare the asymmetries in WALLABY Pilot Survey detections to those found in SIMBA.  Section \ref{ssec:wallaby_asym} describes the asymmetries of the WALLABY detections and Section \ref{ssec:WALLABY_Simba_Comp} contains the comparison of the two.

\subsection{WALLABY Asymmetries}
\label{ssec:wallaby_asym}

One key difference between WALLABY Pilot Survey detections and the mock observations examined in Section \ref{ssec:SIMBA_A} is noise.  Noise tends to increase the asymmetry, but \citet{A3D} developed a correction to the squared difference asymmetry used here:
\begin{equation}\label{Eq:Asym_Back}
    A=\left(\frac{P-B}{Q-B}\right)^{1/2}~,
\end{equation}
where $P$ and $Q$ are the numerator and denominator of Equation \ref{Eq:Asym} respectively, as given by Equation~\ref{eq:AsymParts},  and $B$ is the background correction.  Assuming Gaussian noise, $B= 2 N \sigma^{2}$ where $N$ is the number of voxels/pixels/channels used in the asymmetry calculation, and $\sigma$ is the RMS noise level \citep{A3D}. This correction can still fail for very low $S/N$ detections.  In some cases the signal and noise can be such that $P<B$. When $P<B$ we set $A=-1$ as an indication that there is no measured intrinsic asymmetry, but rather that the full signal from $P/Q$ in Equation \ref{Eq:Asym} can be attributed to the noise. This correction is more likely to fail for $A_{3\mathrm{D}}$ than for $A_{1\mathrm{D}}$ because the signal is integrated before computing the spectral asymmetry.

Secondly, in real observations the minimum of the potential is unknown, so a different center must be used in the asymmetry calculation relative to that adopted in Section \ref{ssec:SIMBA_A}.  For simplicity we use the center found by the WALLABY source finding. WALLABY uses \textsc{SoFiA-X} \citep{WALLABY2022}, which is a parallelized version of \sofia\ \citep{Serra2015,Westmeier2021} designed for the full WALLABY observations.

While WALLABY has detected the \hi\ content of 2419 galaxies in the Pilot Survey phase, not all of these are suitable for the study of asymmetries.  \citet{A3D} noted that measuring 3D asymmetry reliably requires the galaxy be resolved by at least 4 resolution elements.  This requirement matches the 505/2419 detections for which kinematic modelling has been attempted \citep{WALLABY2022B,WallabyPDRPhase2}.  \citet{WALLABY2022B} set the modelling criteria of those galaxies with a \sofia\ ell\_maj$>2$ beams or the integrated $\log_{10}(S/N)>1.25$. $S/N$ is integrated across the detection \citep{Westmeier2021}, and ell\_maj is a measure of the size along the major axis that corresponds to $\sim$half the diameter of kinematically modelled disks \citep{WALLABY2022B}. 
Moreover, since the kinematic models applied assume axisymmetry, one might expect the success of the kinematic models to be anti-correlated with the asymmetries.

Most WALLABY detections have $\textrm{ell\_maj}~<7$ beams.  Based on the results of \citet{A3D} and \citet{Giese2016}, these detections are too poorly resolved for calculations of the spatial (2D) asymmetries, despite them being resolved enough for reliable 3D asymmetries.  
We therefore calculate both 1D and 3D asymmetries for all WALLABY galaxies where kinematic modelling was attempted.  The most important difference between the two is that the background correction of the 1D measures uses the noise in the masked spectra rather than the noise in the cube.  Both quantities are calculated using the \textsc{3DACS} code released in \citet{A3D}.

\begin{table*}[htbp]
    \centering
    \begin{tabular}{|c|c|}
     \hline
       Sample & Number \\
      \hline
       Total WALLABY Detections, Pilot Phase 1 \& 2  & 2419 \\
       \hline
       Kinematic Modelling Attempted  &  505 \\
       Modelling Attempted and $A_{3D}\ge0$ & 158\\
       Successfully Modelled and $A_{3D}\ge0$ & 84\\
       Model Failure and $A_{3D}\ge0$ & 74\\
       \hline
       Modelling Attempted and $\log(M/\Msol)\ge 9.2$ & 384\\
       Modelling Attempted, $A_{3D}>0$, and $\log(M/\Msol)\ge 9.2$& 116 \\
       Successfully Modelled, $A_{3D}>0$, and $\log(M/\Msol)\ge 9.2$ & 67\\
       Model Failure, $A_{3D}>0$, and $\log(M/\Msol)\ge 9.2$ & 49\\
       \hline 
    \end{tabular}
    \caption{Summary of the WALLABY detections for which 3D asymmetries were measured.}
    \label{tab:WALLABYStats}
\end{table*}

Using the procedure described above, we recover 1D and 3D asymmetries of the 116/384 ($\sim 30\%$) Pilot Survey detections for which kinematic models were attempted with $\log_{10}(M_{\hi}/\Msol)\ge9.2$. The first rows of Table \ref{tab:WALLABYStats} summarize the number of WALLABY detections, the number for which kinematic modelling was attempted, and the set of those with measured asymmetries (ie. for which $P<B$ in Equation~\ref{Eq:Asym_Back}). 

\begin{figure}[htbp]
    \centering
    \includegraphics[width=0.95 \columnwidth]{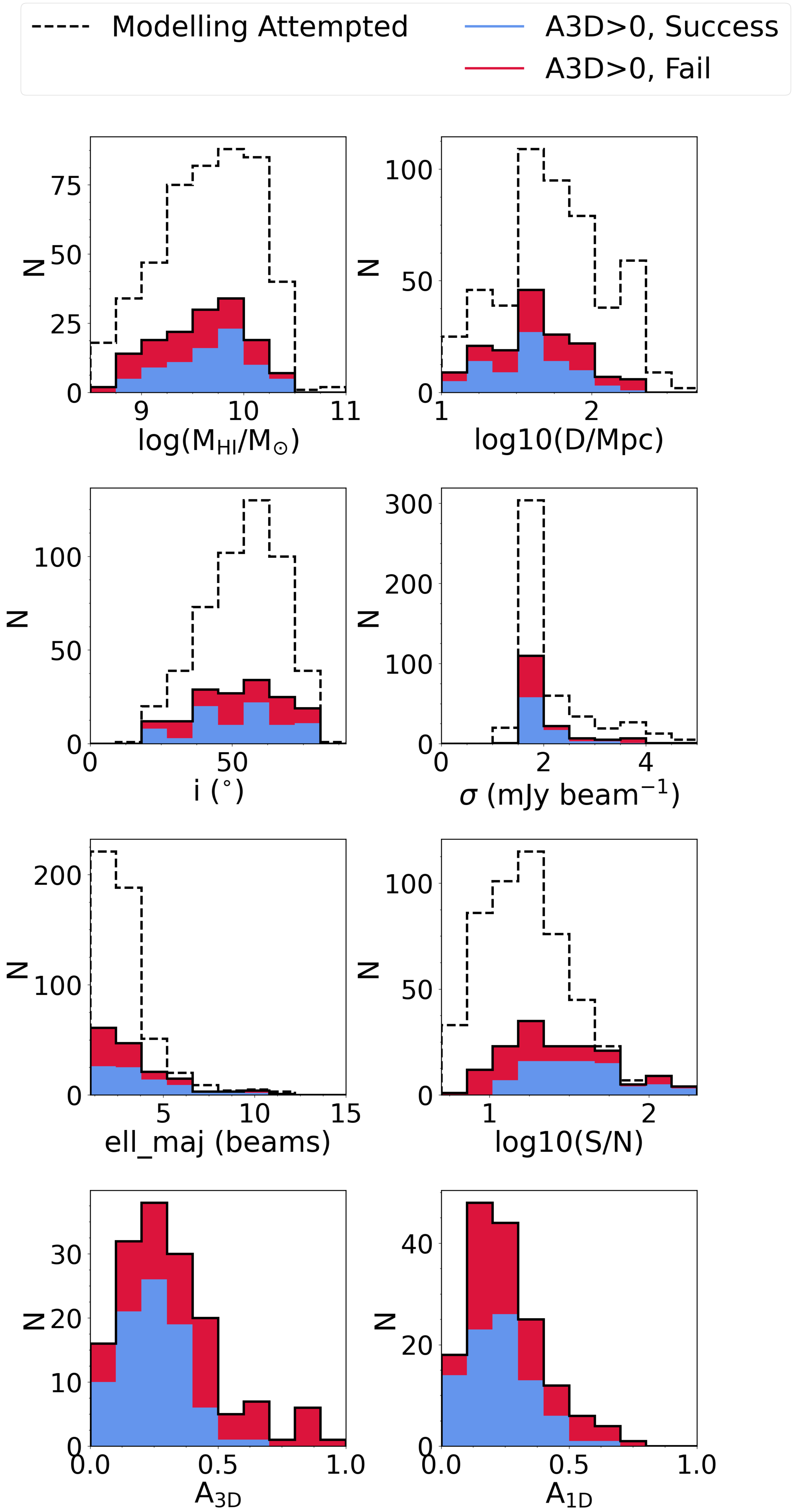}
    \caption{Histograms of the properties of WALLABY detections. The dashed black lines show the full set of detections where kinematic modelling was attempted.  The blue shaded regions show those detections where the kinematic modelling was successful and have measured asymmetries, while the red shaded region shows those where the kinematic modelling failed but still have measured asymmetries.  In the top three rows, the ratio of the red plus blue histograms (the solid black line) to dashed histogram is equal to the fraction of galaxies for which asymmetries were measured.}
    \label{fig:WALLABY_Properties}
\end{figure}

For each WALLABY detection there are a number of properties that are necessary to consider for comparisons to SIMBA.  These include the \hi\ mass $M_{\hi}$, distance $D$, inclination $i$, RMS noise $\sigma$ and size on the sky ell\_maj.  We take $M_{\hi}$ and $D$ from \citet{WALLABYscaling}, $\sigma$ is directly calculated in \textsc{3DACS}, and ell\_maj is from the WALLABY releases \citep{WALLABY2022,WallabyPDRPhase2}. With $\sigma$ and masks within which the asymmetries are calculated, $\log_{10}(S/N)$ can be computed as well.  Figure \ref{fig:WALLABY_Properties} summarizes these properties along with 3D and 1D asymmetries for the full sample of detections where kinematic modelling was attempted as well as the subsamples of kinematically modelled galaxies and those where the modelling failed.

A few things are apparent when examining the histograms in Figure \ref{fig:WALLABY_Properties}.  The successfully modelled galaxies and those where the modelling failed have similar distributions of $M_{\hi}$, $D$, $i$, and $\sigma$.  There is, however, a difference between $D$ for detections where kinematic modelling was attempted and $D$ for detections with measured asymmetries.  The larger sample reaches to higher distances, which suggests that those galaxies, which are smaller on the sky, end up have an unmeasurable 3D asymmetry owing to their smaller size.  The more interesting difference is that the kinematically modelled subsample does not extend to as high 3D or 1D asymmetries as the ones where the modelling failed.   
But the existence of unmodelled galaxies with low asymmetries as well as modelled galaxies with high asymmetries indicates that asymmetry alone is not a predictor of the success of the kinematic modelling.

\begin{figure}[htbp]
    \centering
    \includegraphics[width=0.95 \columnwidth]{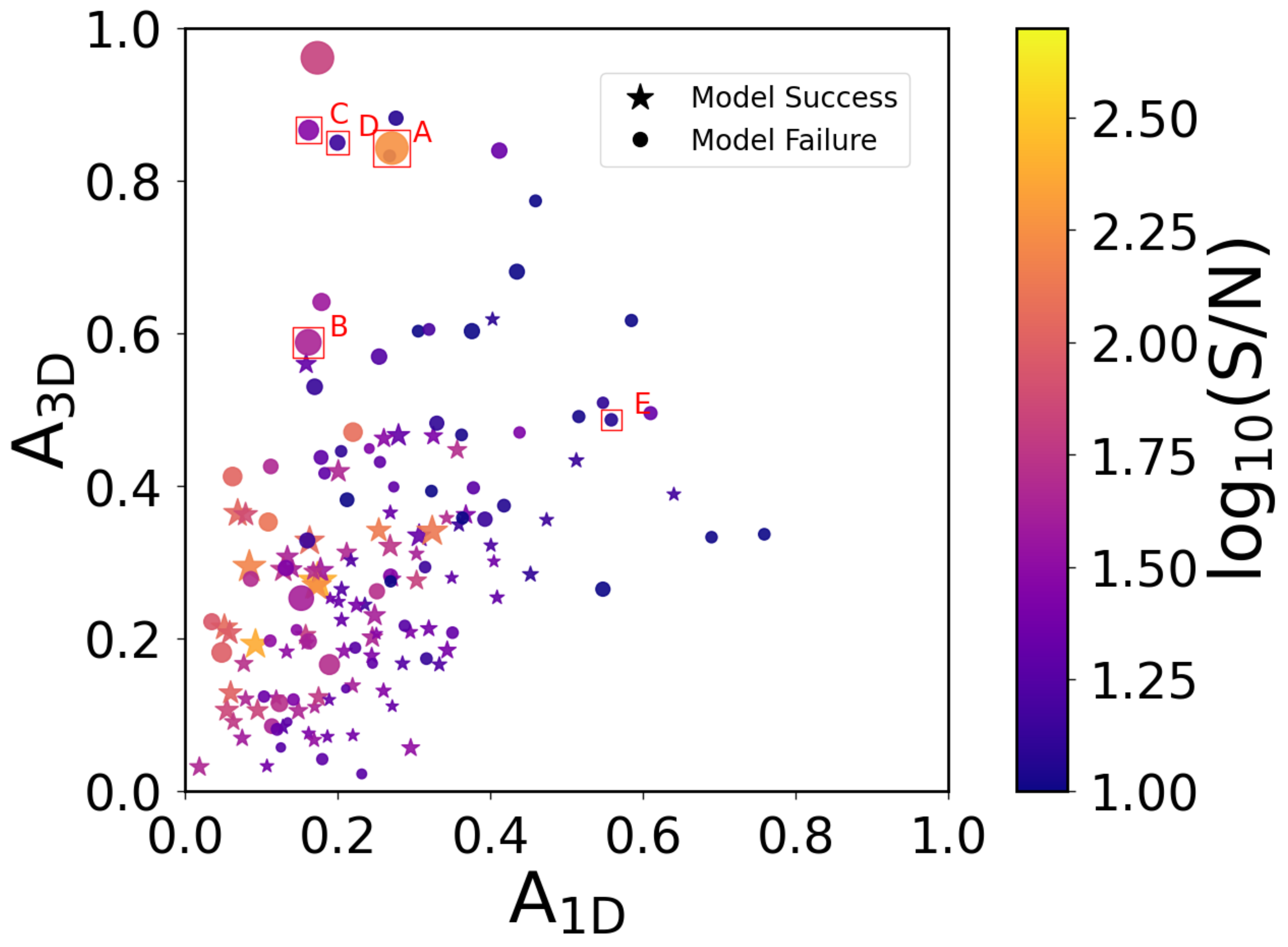}
    \caption{Comparison of the 1D and 3D asymmetries of WALLABY detections with $\log_{10}(M_{\hi}/\Msol)\ge 9$.  The size of the points correlates to their size on the sky, and the colorbar identifies their $S/N$. The stars show galaxies where kinematic modelling is successful, while the circles show those without kinematic models.  Maps and spectra for the detections labelled A, B, C, D, and E are shown in Figure \ref{fig:HighAsymExamples}.}
    \label{fig:WALLABY_Asyms}
\end{figure}

To compare WALLABY to SIMBA, it is necessary for the mock observations to be reliable, which is why we require $\log_{10}(M_{\hi}/\Msol)\ge9$ among those galaxies (see Sec. \ref{ssec:selection}).  To facilitate such a comparison, a mass cut of $\log_{10}(M_{\hi}/\Msol)\ge9.2$ is applied to the WALLABY sample to produce the desired mass limit when coupled with the matching method discussed in Sec. \ref{ssec:WALLABY_Simba_Comp}.  The number of galaxies in this mass cut subsample is listed in the bottom rows of Table \ref{tab:WALLABYStats}, and Figure \ref{fig:WALLABY_Asyms} shows the relationship between the 3D and 1D asymmetries for this population.  

A number of key properties of the WALLABY sample are apparent in Figure \ref{fig:WALLABY_Asyms}.  Firstly, as suggested by the bottom row of Figure~\ref{fig:WALLABY_Properties}, the successfully modelled detections and those where the modelling failed have different asymmetry distributions.  The kinematically modelled galaxies tend to have lower asymmetries, and there are very few with `extreme' asymmetries in either 1D or 3D, where $A\ge 0.5$.  This is a statistical trend, but, as with Figure \ref{fig:WALLABY_Properties}, this plot shows that, on an individual galaxy basis, the specific measured asymmetry does not indicate whether a galaxy is kinematically modelable.  At high $S/N$, the galaxies tend to be larger, which is expected as $S/N$ correlates with galaxy size \citep{WALLABY2022,WALLABY2022B}.  More importantly, high $S/N$ detections tend to have lower $A_{1\mathrm{D}}$ but a full range of $A_{3\mathrm{D}}$, in line with the results for the noiseless SIMBA mocks in Section~\ref{ssec:SIMBA_A}, further underscoring the power of 3D asymmetries for identifying disturbed systems.   

\begin{figure}[htbp]
    \centering
    \includegraphics[width=0.95 \columnwidth]{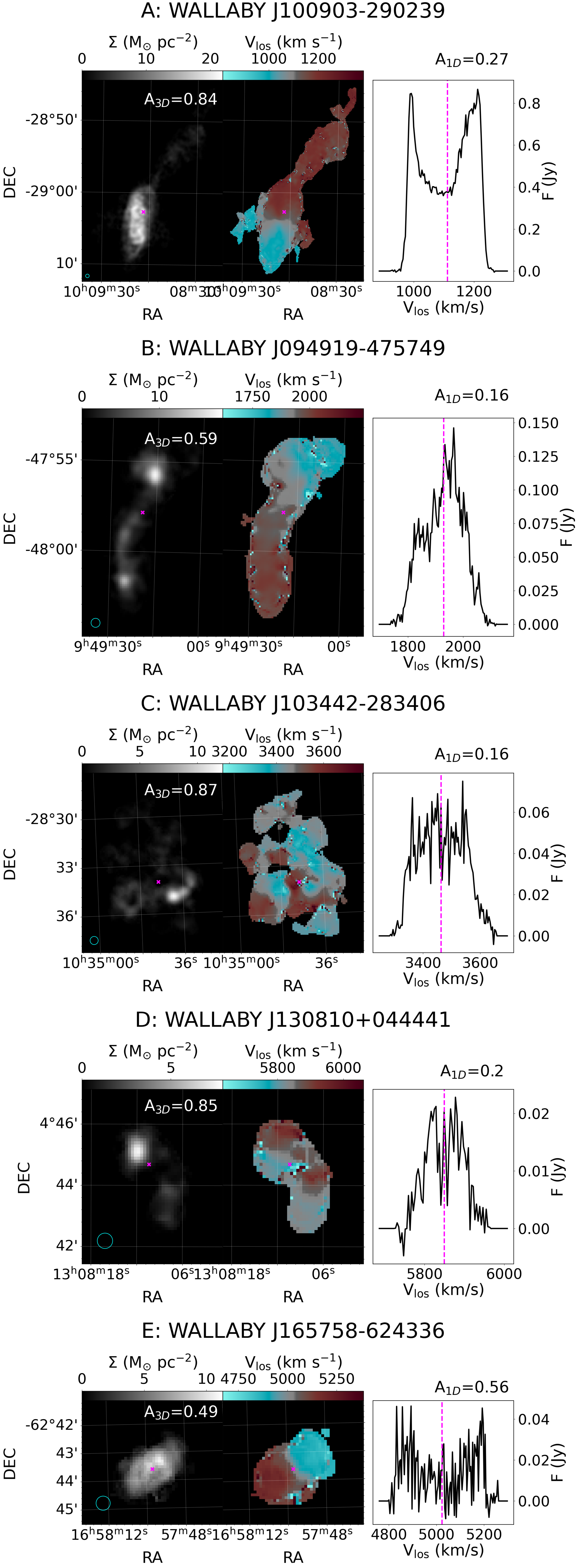}
    \caption{Moment maps and spectra for a set of detections with extreme asymmetries, labeled A--E in Figure \ref{fig:WALLABY_Asyms} to illustrate different drivers of asymmetry.  The cyan circles in the lower left of the moment 0 map panels show the size of the beam. The magenta x's and vertical line in the maps and spectral plots show the \sofia\ center.  This center is used for the calculation of the asymmetries listed in the moment 0 plot (3D) and spectra (1D).}
    \label{fig:HighAsymExamples}
\end{figure}

Examples of systems with high $A_{1\mathrm{D}}$ and/or $A_{3\mathrm{D}}$ are shown in Figure \ref{fig:HighAsymExamples} for illustrative purposes.  The most common drivers for extreme 3D asymmetries in well-resolved detections are a) extended tidal features that cause the \sofia\ center point to be away from the expected kinematic center like in WALLABY J100903-290239 (row A of Figure \ref{fig:HighAsymExamples}), and b) galaxy pairs connected by large tidal bridges as in WALLABY J094919-475749 (row B of Figure \ref{fig:HighAsymExamples}).  In both of these cases, it may be possible to kinematically model the main galaxy or galaxies.  In the case of WALLABY J103442-283406 (row C of Figure \ref{fig:HighAsymExamples}), the detection is an entire interacting group that was examined in detail by \citet{OBeirne2024}.

At more moderate resolutions, the extreme asymmetries are more often caused by pairs of galaxies being close enough together that \sofia\ identified them as a single source like WALLABY J130810+044441 (row D of Figure \ref{fig:HighAsymExamples}).  There are also detections like WALLABY J165758-624336 (row E of Figure \ref{fig:HighAsymExamples}) with high $A_{1\mathrm{D}}$ but moderate $A_{3\mathrm{D}}$.  These tend to be small galaxies with low $S/N$ where the background correction is likely underestimated for the 1D asymmetries, leading to their larger $A_{1\mathrm{D}}$.  In rows A-C of Figure \ref{fig:HighAsymExamples}, $A_{1\mathrm{D}}$ is not particularly high despite the large disturbances to the detections, highlighting the power of 3D asymmetries at identifying strongly disrupted objects.  It appears that, for the most asymmetric detections (those with $A_{3D}>0.5)$, it is gas that lies beyond the main disk (either in tidal features or other galaxies) that drives the asymmetry.  Therefore, selecting detections with $A_{3D}>0.5$ will reliably find systems with such extended features at the WALLABY resolution and sensitivity.

\subsection{SIMBA and WALLABY}\label{ssec:WALLABY_Simba_Comp}

To compare SIMBA to WALLABY, it is necessary to generate a WALLABY-like SIMBA mocks with similar properties to the WALLABY sample.  Our mock SIMBA sample must have a similar mass, distance, inclination, and noise distribution to those seen in Figure \ref{fig:WALLABY_Properties}.  

To build our mock sample, we select WALLABY detections on which modelling was attempted and then match each one to a SIMBA galaxy with similar properties. In detail, for each WALLABY detection with \hi\ mass $M_{\hi,W}$, a SIMBA galaxy with \hi\ mass $M_{\hi, S}$ in the range $M_{\hi, S}\in \left[M_{\hi,W} 10^{-0.2}, M_{\hi,W} 10^{+0.2} \right]$ is randomly selected from the sample of 789 galaxies that fit the criteria of Section \ref{ssec:SIMBA}.  A mock \hi\ cube is made of that SIMBA galaxy using our Scanline Tracing MFM method, where the galaxy is placed at the WALLABY detection's recovered distance and inclination.  Gaussian noise is then added and convolved by the $30\arcsec$ WALLABY beam such that the resulting noisy cube has the same noise RMS as the WALLABY detection. Finally, we measure $A_{1\mathrm{D}}$ and $A_{3\mathrm{D}}$ of the paired SIMBA detection in the same fashion as the WALLABY detection, running \sofia\ on the noisy mock and adopting the resulting \sofia\ mask and center point in \textsc{3DACS}.

It is possible to increase the matched sample by creating multiple SIMBA realizations of the WALLABY galaxies.  In each realization, one SIMBA galaxy is generated for each WALLABY galaxy.  Ultimately we generate 20 realizations, providing 20x more mock observations than WALLABY detections.  While a specific SIMBA galaxy will appear multiple times in the full set of realizations, it will have varying distances, viewing angles, and orientations relative to the observer each time, which produce different measured asymmetries.

As with WALLABY, it is possible for the mock WALLABY observations to have noise such that $P<B$ in Equation~\ref{Eq:Asym_Back}, and no intrinsic measure of the asymmetry is recovered. As noted in Section \ref{ssec:wallaby_asym}, only 30\% of WALLABY detections for which kinematic modelling was attempted have a measured asymmetry.  For our mock WALLABY-like SIMBA cubes, the asymmetries are measured for 23\% of the mocks. This is slightly lower than the WALLABY fraction,  but not inconsistent with it given the relatively small sample sizes in question.

We quantitatively compare the distribution of measured WALLABY and SIMBA asymmetries using the PQMass test \citep{PQM}.  The PQMass test performs a Pearson chi-square test by binning our chosen parameter space of ($A_{1\mathrm{D}}$,$A_{3\mathrm{D}}$) using a Voronoi tessellation for a set of points of the SIMBA mock. We elect to use 10 bins. The PQMass test recovers that the probability that SIMBA and WALLABY asymmetry samples are drawn from the same distribution is $p \sim 5.4\pm0.2\%$, where the final value is calculated as an average from $\sim1000$ runs of the test which vary through different Voronoi binning. The uncertainty is derived from the standard deviation across the runs.  While relatively low, a $\sim 5\%$ probability is sufficiently large that we cannot conclude that the SIMBA and WALLABY asymmetry distributions are inconsistent with each other.  To make such a conclusion, we will require a significantly larger sample of WALLABY observations as well as simulated galaxies.

\begin{figure}[htbp]
    \centering
    \includegraphics{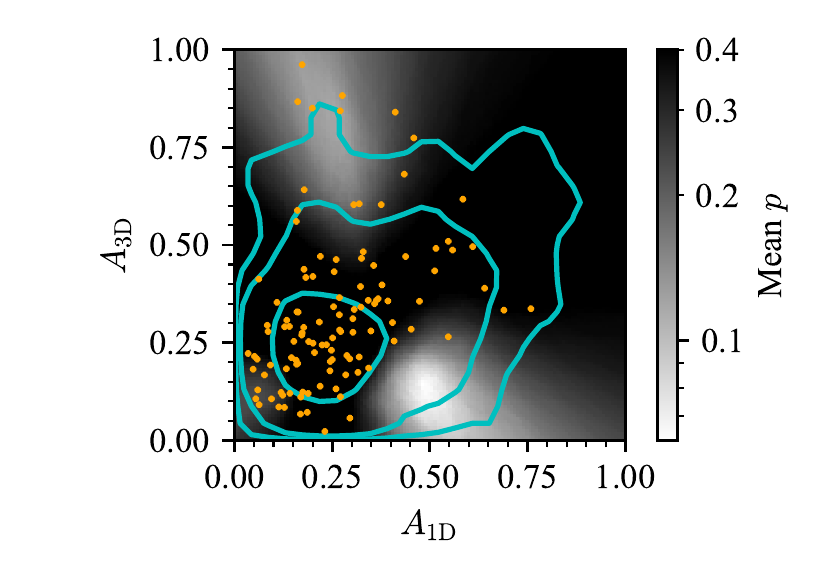}
    \caption{Map of the asymmetry distributions for the WALLABY sample and the WALLABY-like SIMBA mocks. The cyan contours draw the SIMBA distribution for the count-wise 68th, 95th and 99th percentile, the orange dots are the individual WALLABY measures. The underlying map shows the mean $p$ value for binomial tests averaged across different binnings of the $A_{1\mathrm{D}}$ -- $A_{3\mathrm{D}}$ plane.}
    \label{fig:pqm}
\end{figure}

Figure \ref{fig:pqm} overlays the SIMBA asymmetry distribution over the WALLABY detections as contours in order to more closely examine the similarities and differences between the two distributions.  The grayscale colormap in  Figure \ref{fig:pqm} shows the $p$-value for the null hypothesis that the asymmetries sample a given bin at the same rate. This $p$-value should not be taken as a rigorous measure, but rather it is intended to illustrate the discrepancy between the SIMBA and WALLABY samples: it is calculated using the binomial test for the different PQMass Voronoi bins, averaged across multiple binnings. 

Figure \ref{fig:pqm} illustrates that there are two regions in the $A_{1\mathrm{D}}$ -- $A_{3\mathrm{D}}$ where the WALLABY and SIMBA distributions appear to differ, which drive the final PQMass value of $\sim5\%$.  Firstly, the high $A_{3\mathrm{D}}$ low $A_{1\mathrm{D}}$ corner shows a significant difference driven by an overdensity of WALLABY detections.  These detections, as discussed in Section \ref{ssec:wallaby_asym} and illustrated in Figure \ref{fig:HighAsymExamples}, mostly consist of interacting galaxies where the \sofia\ center is drawn away from the principal galaxy, increasing the spatial asymmetry.  While this same behavior can happen in SIMBA, it occurs for less than 1\% of SIMBA mocks as indicated by the 99\% contour limit.  This hints at a possible difference between WALLABY and SIMBA, but there are other possible causes of this difference, such as environmental selection effects.

The second region where a discrepancy is noticeable is in the moderate $A_{1\mathrm{D}}$ low $A_{3\mathrm{D}}$ region. This region is overpopulated by the SIMBA mocks compared to WALLABY detections. A possible explanation is that the SIMBA mocks are under-resolved relative to their WALLABY counterparts, which could occur, for example, if the SIMBA disks are smaller than the WALLABY ones at a given $M_{HI}$ (such that the spatial contribution to $A_{3\mathrm{D}}$, and therefore $A_{3\mathrm{D}}$ itself, is too low). Overall, however, these differences are small, and we leave a detailed investigation of the causes for these differences to future work.

\section{Discussion and Conclusions}
\label{sec:conclusion}

We have presented a Scanline Tracing method to create  mock \hi\ datacubes for simulated galaxies, reducing the shot noise along the spectral axis of the mock which makes a spurious contribution to asymmetry measures. We apply this technique to generate WALLABY survey-like \hi\ datacubes of simulated galaxies from the SIMBA 50 Mpc simulations, and compare 1D and 3D asymmetries therein. We find hints of an excess of high-asymmetry systems in WALLABY compared to the SIMBA mocks, though that difference is not statistically significant as measured by a PQMass test of the distribution of points in $A_{1\mathrm{D}}$ and $A_{3\mathrm{D}}$.  

The Scanline Tracing method developed in Section \ref{ssec:SIMBA} can be used to create mock $\hi$ cubes for both SPH and MFM hydrodynamical simulations. While this approach introduces additional operations in the computation of the different fields, it avoids the shot noise introduced by the particle-centric method adopted by the default MARTINI settings, as shown in Figures \ref{fig:gal467} and \ref{fig:gal1531}. This advantage is considerable for the study of asymmetry in marginally resolved simulations. However, as the number of particles increases, it is expected that the Particle method would converge to the Scanline Tracing method while offering greater computational ease.  For cooler gas \citep[see][]{Ploeckinger}, the resolution necessary for the particle approach is likely to increase. Currently, for asymmetries in the $50\ \mathrm{Mpc}$ SIMBA simulation, the Scanline Tracing method shows noticeable differences with the MARTINI default particle-centric method as seen in Figure \ref{fig:histVSA3D}. Unlike the Particle method, Scanline Tracing will, in the lower resolutions, have the propensity to over-smooth spectra. This error is less evident than the Particle method's shot noise which may prove more perverse. 

In Section~\ref{ssec:wallaby_asym}, we measured $A_{1\mathrm{D}}$ and $A_{3\mathrm{D}}$ asymmetries for the sample of WALLABY pilot detections for which kinematic modelling was attempted. As shown in Figure~\ref{fig:WALLABY_Asyms}, detections where the kinematic modelling was successful had fewer extreme asymmetries than those where the modelling failed. However, there is a large scatter and the correlation isn't statistically significant. Upon inspection (Figure~\ref{fig:HighAsymExamples}), the WALLABY detections with relatively high signal to noise and $A_{3\mathrm{D}}\gtrsim0.5$ tend to be interacting systems or systems with extended tail-like features. Asymmetry may be an efficient way to detect these systems using \hi\ data alone, which is particularly useful in regions where multi-wavelength data are scarce, such as the Zone of Avoidance.

Figure~\ref{fig:pqm} illustrates that our mock WALLABY-like SIMBA samples show signs of discrepancy in their $A_{1\mathrm{D}}$ and $A_{3\mathrm{D}}$ when compared to WALLABY, even when controlled for mass, distance and noise. This may hint at potential deviations between the SIMBA galaxy evolution model and the \hi\ distributions in WALLABY detections. Although the PQMass $p$-value of $\sim$5\% indicates that these deviations are not statistically significant on the whole, we can nonetheless speculate on the potential drivers of lower asymmetries in the SIMBA galaxies compared to observations. 
For example, the high $A_{3\mathrm{D}}$ low $A_{1\mathrm{D}}$ corner where WALLABY detections out-populate the SIMBA  mocks may be caused by the overly hot SIMBA IGM \citep{Dave2017, wright24}, which might inhibit structures like \hi\ bridges and tidal features that drive the WALLABY asymmetries. Another possibility is that the WALLABY sample is biased to high asymmetries, perhaps owing to the group and cluster environments preferentially probed by the Pilot Survey fields \citep{WALLABY2022,WallabyPDRPhase2}.

Although our analysis is limited by the relatively low number of \hi\ detections in the WALLABY Pilot Survey detections and the relatively modest particle masses and resolutions in the SIMBA 50 Mpc simulation, the quantitative comparisons between asymmetries in \hi\ detections and in simulations presented in this paper are among the first of their kind, and a prelude of things to come in this space. More thorough comparisons will be enabled with the full statistical power of WALLABY as the complete survey is released. It will also be possible to compare different subgrid physics and dark matter models, both within simulations such as SIMBA as well as across different simulations, particularly as the latter increase in volume and in resolution. This opens new avenues for comparing simulations with observations to determine how various physical phenomena drive \hi\ morphological asymmetries. 

\section{Acknowledgements}
The Australian SKA Pathfinder is part of the Australia Telescope National Facility which is managed by CSIRO. Operation of ASKAP is funded by the Australian Government with support from the National Collaborative Research Infrastructure Strategy. ASKAP uses the resources of the Pawsey Supercomputing Centre. Establishment of ASKAP, the Murchison Radio-astronomy Observatory and the Pawsey Supercomputing Centre are initiatives of the Australian Government, with support from the Government of Western Australia and the Science and Industry Endowment Fund. We acknowledge the Wajarri Yamatji people as the traditional owners of the Observatory site. WALLABY acknowledges technical support from the Australian SKA Regional Centre (AusSRC).

Parts of this research were supported by the Australian Research Council Centre of Excellence for All Sky Astrophysics in 3 Dimensions (ASTRO 3D), through project number CE170100013.

We acknowledge the use of the ilifu cloud computing facility - www.ilifu.ac.za, a partnership between the University of Cape Town, the University of the Western Cape, Stellenbosch University, Sol Plaatje University, the Cape Peninsula University of Technology and the South African Radio Astronomy Observatory. The ilifu facility is supported by contributions from the Inter-University Institute for Data Intensive Astronomy (IDIA - a partnership between the University of Cape Town, the University of Pretoria and the University of the Western Cape), the Computational Biology division at UCT and the Data Intensive Research Initiative of South Africa (DIRISA).

MG is supported by the Australian Government through the Australian Research Council’s Discovery Projects funding scheme (DP210102103), and through UK STFC Grant ST/Y001117/1. MG acknowledges support from the Inter-University Institute for Data Intensive Astronomy (IDIA). IDIA is a partnership of the University of Cape Town, the University of Pretoria and the University of the Western Cape. For the purpose of open access, the author has applied a Creative Commons Attribution (CC BY) licence to any Author Accepted Manuscript version arising from this submission. This research was undertaken thanks in part to funding from the Canada First Research Excellence Fund through the Arthur B. McDonald Canadian Astroparticle Physics Research Institute. KS acknowledges support from the Natural Sciences and Engineering Research Council of Canada (NSERC). KAO acknowledges support by the Royal Society through a Dorothy Hodgkin Fellowship (DHF/R1/231105).

This work has made use of NASA's Astrophysics Data System.

\FloatBarrier
\bibliography{refs, bibSPH, bibMartini, bibWALLABY}

\end{document}